\documentclass{article}

\usepackage[utf8]{inputenc} 
\usepackage[T1]{fontenc}    
\usepackage{url}            
\usepackage{booktabs}       
\usepackage{amsfonts}       
\usepackage{nicefrac}       
\usepackage{microtype}      
\usepackage{xcolor}         

\title{Expose Uncertainty, Instill Distrust, Avoid Explanations: Towards Ethical Guidelines for AI}

\author{%
  Claudio S. Pinhanez \\
  IBM Research - Brazil\\
  Rua Tutoia 1157 - S\~ao Paulo - SP - Brazil \\
  \texttt{csantosp@br.ibm.com} \\
}

\begin{document}

\maketitle

\begin{abstract}
  In this position paper, I argue that the best way to help and protect humans using AI technology is to make them aware of the intrinsic limitations and problems of AI algorithms. To accomplish this, I suggest three ethical guidelines to be used in the presentation of results, mandating AI systems to expose uncertainty, to instill distrust, and, contrary to traditional views, to avoid explanations. The paper does a preliminary discussion of the guidelines and provides some arguments for their adoption, aiming to start a debate in the community about AI ethics in practice.
\end{abstract}

\section{Machine Computation of Similarity}

The goal of this position paper is to explore whether and how some intrinsic limitations of modern \emph{Artificial Intelligence (AI)} technology should lead to ethical guidelines mandating that the presentation of their results must: (1)~expose the underlying uncertainty of the results; (2)~foster distrust and doubt about their appropriateness and accuracy; (3)~avoid the use of explanations which further increase user confidence on the results. I argue here for those guidelines as a way to safeguard AI users from mistakenly making decisions based on unfounded beliefs about the accuracy of machines, on biased data, and on other known ailments of current AI technologies~\cite{crawford2021atlas}.

It is first very important to acknowledge how welcomed the debate around ethics in AI is, although, in my view, it is still insufficient in scope. The opacity of the current, successful generation of AI algorithms, based on \emph{neural networks (NN)} technology, seem to have brought to light a fundamental discussion about the use of \emph{information systems (IS)} in the society, and particularly for decision-making processes involving its most vulnerable and discriminated members.

However, the key issues in discussion, such as bias towards individuals and groups, transparency, explainability, and correctness, apply to a much larger group of ISs, including ones based on statistical methods, optimization algorithms, and data science in general~\cite{browne2015dark,eubanks2018automating,crawford2021atlas}. That debate is better done, and served, considering all kinds of ISs which are often used to reproduce, and sometimes hide, questionable or deplorable political and social agendas~\cite{browne2015dark,eubanks2018automating,crawford2021atlas}. 

That point taken, a couple of issues of AI technology seem to have brought, in my opinion, the  spotlight of ethics discussion from ISs to almost exclusively on AI. First, the notorious and inherent opacity of \emph{machine learning (ML)} methods, particular NNs, whose behavior is not comprehensible by human beings, tends to instill, understandably, fear.
Second, the excessive and unfounded propaganda that the current AI technology is, indeed, intelligent in the human sense, trumpeted by the tech industry and its pundits, has made people and society much more concerned about its spread. 

As a researcher in AI for more than 30~years, I can not see how current AI technology can be considered as intelligent, in any human sense. From a technical point of view, ML algorithms, and particularly NN-based ones, are simply mimicking, often inaccurately, the human perception of \emph{similarity}. What a NN algorithms does is to create a complex mathematical function which maps an input pattern into known patterns, provided as labelled examples by its developers. Based on those, the algorithm produces a classification token, the one most similar according to the examples. As brilliantly put by Alan Blackwell~\cite{blackwell2019objective}, \emph{``... ‘intelligent’ behaviour of machines is no more than human behaviour reflected back to us.''} Being able to computer similarity is essential for intelligence, but not sufficient.


So how to appropriately portray and ethically use a opaque, inscrutable technology based on the machine replication, often inaccurately, of the human admirable ability of identifying patterns and perceiving similarity? 
I propose here that to ethically use modern AI algorithms we should follow 3~basic ethical guidelines: 
it is necessary to \textbf{expose uncertainty}, that is, make their users aware of the level of imprecision of their results; \textbf{instill distrust} on their results and predictions, in a systematic and deliberate way; and \textbf{avoid explanations}, which tend to foster unfounded beliefs of true machine intelligence in non-technical users. Albeit explainability is often portrayed as important way to deal with ethical issues, I contend here that is a misguided effort if intended to help the common user navigate modern AI technology.

I am not denying here the extraordinary technical advances in the last 10~years which have substantially increased the ability to machines to more accurately deal with similarity. The early AI of the 1970-1980s was plagued by brittleness and its effects on the rule-based and logic programming systems of that era, impeding its application in real problems. The pendulum has swung, and current AI technology suffers from a chronic softness which has to be better communicated to its users. 


For the rest of the paper, I will use the term \emph{AI} to mean ML-based algorithms and, particularly, NN-based ones. Although this is a somewhat unfair portrayal of the richness of methods created and used by the AI community and profession, it matches the current use of the term by most people, enterprises, societal actors, and media.




\section{Exposing Uncertainty}

In a recent meeting at my laboratory, a question surfaced related to what is the world's \emph{Gross Domestic Product~(GDP)}, or \emph{Gross World Product~(GWP)}. I typed the question in \emph{Google}, using Portuguese language, and got the answer, translated here to English, \emph{``... the total wealth should climb from US\$~226 trillion in 2019...''}. The same question, in English, yielded \emph{``In 2020, global GDP amounted to about 84.54 trillion US dollars,...''}.

The reason for the wild discrepancy was the incorrect matching of the term \emph{total wealth} to \emph{global GDP} in the Portuguese version. Although similar, they are quite different concepts in economics, a distinction which the Portuguese seemed to lack but the English version had. 
The answers, in both cases, were not portrayed as part of a list of search results, which users have learned to take with some care. Those statements were provided by the Google search engine as \emph{the answers} to the question, in spite the of incorrect matching of the terms in Portuguese. 

The designers of the Google search interface could have conveyed the uncertainty of their matching algorithms easily by preceding the answer with a sentence such as \emph{``A related answer to you question may be: ...''}, able to expose, in a meaningful way to humans, the intrinsic uncertainty of the algorithm, and of language itself. However, they did not so, as almost all designers and developers of interfaces which publish results of AI algorithms, especially for non-technical users.

What I question here is the ethics of not ascertaining and making clear the uncertainty of an AI algorithm, which often creates a unsubstantiated claim of correctness which is beyond the technology. In many cases, there seems to be a deliberate intent to hide the uncertainty to increase the perception of quality of the ISs. 
However, omitting the uncertainty takes a completely new dimension when the result of the algorithm has more impacts on people's lives, such as when used to determine a parole claim, the identity of a criminal, the provision of a home loan, or to stop an automatic car from running over a pedestrian. 
To fix this, I propose the following ethical guideline:

\noindent \textbf{Ethical Guideline 1 - Expose Uncertainty}: Interfaces with results of an AI algorithm should properly disclose to its users realistic estimates of the uncertainty of the algorithm, in easy-to-understand forms.

To effectively implement this guideline, it is important to address two key terms: ``realistic estimates'' and ``easy-to-understand''. Estimates of uncertainty are not the same as confidence levels, likelihoods, or even probabilities, in the rare cases where AI algorithms actually compute them. Real uncertainty must consider the errors of the algorithm, the labeling mistakes of the training datasets, and possible new conditions of the moment when the results are presented. For example, AI algorithms used to predict stock prices had its real uncertainty increased when the COVID-19 pandemic started. Similarly, e-commerce recommendations portrayed a lot of nonsensical suggestions after the lock-downs started, since it was the first time they were seeing data of entire populations buying all their needs in the Internet. In practice, factoring in the issues affecting uncertainty is a hard problem, sometimes harder than the one the AI algorithm is trying to solve.

On the other hand, handling the creation of easy-to-understand portrayals of uncertainty has been in the scope of the HCI research community for decades, and should be a key area in Human-Centered AI. For data displays, there is very rich research in Visual Analytics around uncertainty visualization, with a set of well-established and well-researched methods~\cite{griethe2006visualization,brodlie2012review,bonneau2014overview,sacha2015role,hullman2018pursuit}. Similarly, there are many ways to convey uncertainty through language, for instance, by preceding statements with phrases such as \emph{``I~believe...''} or \emph{``I~guess...''}; by using uncertainty adjectives and adverbs such as \emph{``approximate''} and \emph{``about''}; and by using informal language~\cite{heylighendewaele1999formality}. Further investigation is needed whether the use by machines of such terms and methods indeed produces similar effects as when used by people, even considering that some previous work on human perception of computers points towards a positive answer~\cite{reeves1996media,nass2005wired}.


\section{Instilling Distrust}

It is remarkable that the most remembered moment of the famous \emph{Jeopardy} competition of 2011, which inaugurated the current AI era, was the one where the \emph{IBM Watson} computer made a silly mistake. In the category \emph{``U.S. Cities''}, for the prompt \emph{``Its largest airport is named for a World War II hero; its second largest, for a World War II battle''}, Watson answered \emph{``What is Toronto?''}, instead of the right answer \emph{``What is Chicago''}.  
If IBM Watson had lost the match, that mistake would probably have been the definite example of how machines can not be as intelligent as people. The Watson victory framed that error in a more benign view, which nevertheless highlights how different, in fact, the reasoning processes between  AI technologies and people are.

Silly errors have been reported in every milestone of AI, as in the more recent case of \emph{OpenAi GPT-3}. Asked, \emph{``How many eyes does a blade of grass have?''}, GPT-3 answered \emph{``A blade of grass has one eye.''} At the core of the issue is the reliance on similarity by the technology behind the text generator. Being able to produce text recognized as natural language is a far cry from being able to think or even to use language properly. In fact, recent work has shown that SOTA transformer-based models such as BERT have issues to deal with even basic syntax usage, including word order~\cite{sinha2020unnatural}.

Albeit Watson and GPT-3 are both examples of the amazing advances in AI in the last years, they are also a testament of their limitations, showcased in their famous mistakes. I argue here that, in fact, the open nature of their errors is in fact positive, because it instilled in the public the distrust that both systems fairly deserve, as systems which we could hardly classify as intelligent. Unfortunately, the users of the majority of AI systems do not see their systems failing, and often trust them beyond what they can in fact deliver. More worryingly, developers and entrepreneurs often work hard to hide mistakes and overhype the AI systems they build. Letting people make decisions with wrong or inadequate knowledge about the correctness of systems they use is wrong and dangerous. This suggests a second guideline:

\noindent \textbf{Ethical Guideline 2 - Instill Distrust}: Interfaces displaying results of an AI algorithm should include mechanisms to foster the user to question, doubt, and distrust the results and the overall capabilities of the algorithm.

As the examples above suggest, one of  the simplest ways to create distrust in machines seems to be by exposing their errors. In fact, a study on \emph{algorithmic aversion}~\cite{dietvorst2015algorithm} shows that people, after seeing a machine making errors, prefer a human counterpart even when given evidence that the machine's performance is better. Showing and acknowledging mistakes is a practical way to implement this guideline, and the research in the area has explored many ways to do so~\cite{griethe2006visualization,han2020beyond,chiou2021trusting,mayr2019trust}.

There is, however, much more research needed on how to inspire distrust and doubt in machine results, especially when they are part of a larger context. The research on countering beliefs in \emph{deepfakes} has showcased how difficult, in practice, is to raise doubt in people when the information provider does not want to~\cite{Boyd2017DidML,Lazer1094,pennycook2020fighting,hameleers2020picture,schwarz2016making}. Having an ethical guideline mandating the fostering of distrust seems to be needed, as a way to discourage willful omission.

\section{Explaining AI Algorithms Considered Harmful}

An enormous body of work has been created in the last years to develop methods to explain the workings of AI algorithms~\cite{adadi2018peeking,ferreira2020people,xu2019explainable,dovsilovic2018explainable}. However, at best, explainability will help users understand how an AI algorithm is computing its similarities. But, even for human beings, are explanations about why things are similar useful? If people are asked very basic similarity questions, such as why two objects seem to have the same color, they often can not explain or, worse, provide technical explanations (such as color spectrum) which has little to do with their actual perception of similarity.

Moreover, AI algorithms are often trained with datasets built by crowdsourced workers~\cite{crawford2021atlas}, therefore containing the workers own perceptual mistakes and bias of how sentences or images are similar, or how good an answer to a question is. Those workers, most often than not, do not rely on systematic reasons to make their choice: for speed, they often answer based on perception. Similarly, when professional data is used, such is in medical diagnosis, the reasoning behind those choices is often a combination of ill-defined professional rules and the perceived similarity to cases from the past.

The key argument is that the replication of human perception of similarity, the greatest achievement of current AI technology, mimics a process which human beings can hardly explain, using data which is not explainable itself. This poses the question of how can then a machine explain its results besides stating that it is replicating a non-explainable process of human beings? 

Besides that, the only explanations which can be deduced from most AI algorithms are of numeric nature, such as an input is close in a very complex space to a certain group of examples; or that there is a correlation between inputs, examples, and outputs. After the debacle of using scientific data to communicate to people basic facts in the COVID-19 pandemic~\cite{pennycook2020fighting}, it is hard to imagine that using probabilities and correlations is likely to work to make people understand AI systems which use much more complex and sophisticated mathematical methods. Explaining the workings of an AI system to help users understand its limitations and inaccuracies seems to be a doomed objective.

Moreover, providing explanations to users, in fact, seem to go against the two ethical guidelines proposed above. Reasonably phrased explanations are likely to decrease the perception of uncertainty and increase trust. This is summarized in the following guideline:

\noindent \textbf{Ethical Guideline 3 - Avoid Explanations}: Interfaces with results of an AI algorithm should not provide explanations which increase the confidence or trust of users on their results.

It is important to acknowledge that explainability has an important value for professional use, when creating, developing, and approving systems. Similarly, better methods to understand the workings of AI algorithms can be very important for auditing, regulating, and certifying AI systems. But I believe it is wrong to expect that the kind of explanations that can be derived from AI systems will help users of AI to understand its limitations and when to believe or not in AI-produced results. 

\section{Towards a Trustworthy Use of AI}

The basic proposal of this paper is that \textbf{the best way to make AI trustworthy is to act to decrease the users trust in it.} I argue that, since modern AI algorithms are essentially human-behavior replicating machines~\cite{blackwell2019objective}, trained on noisy human-similarity data~\cite{kelleher2009conversational,crawford2021atlas}, they are doomed to produce uncertain, incorrect, and inexplicable results. Therefore, the ethical way to present their results is to highlight those characteristics, what was captured in the three proposed ethical guidelines: \textbf{expose uncertainty, instill distrust, and avoid explanations.}

This is certainly a provocative and controversial statement, which also includes many open research questions, such as the techniques and methods to implement the guidelines, and how human beings will actually perceive and deal with issues of uncertainty and trust in information from machines~\cite{dietvorst2015algorithm}. My hope with this paper is to present those ideas to provoke and foster a discussion in the community with the ultimate goal of producing ethical guidelines for the use of AI.

\bibliographystyle{plain}

\begin{thebibliography}{10}

\bibitem{adadi2018peeking}
Amina Adadi and Mohammed Berrada.
\newblock Peeking inside the black-box: a survey on explainable artificial
  intelligence (xai).
\newblock {\em IEEE access}, 6:52138--52160, 2018.

\bibitem{blackwell2019objective}
Alan~F Blackwell.
\newblock Objective functions:(in)humanity and inequity in artificial
  intelligence.
\newblock In Geoffrey E.~R. Lloyd and Aparecida Vilaça, editors, {\em Science
  in the Forest, Science in the Past}, chapter~9, pages 191--208. Hau Books,
  Chicago,IL, 2020.

\bibitem{bonneau2014overview}
Georges-Pierre Bonneau, Hans-Christian Hege, Chris~R Johnson, Manuel~M
  Oliveira, Kristin Potter, Penny Rheingans, and Thomas Schultz.
\newblock Overview and state-of-the-art of uncertainty visualization.
\newblock In {\em Scientific Visualization}, pages 3--27. Springer, 2014.

\bibitem{Boyd2017DidML}
Danah Boyd.
\newblock Did media literacy backfire?
\newblock {\em Journal of Applied Youth Studies}, 1(4):83--89, 2017.

\bibitem{brodlie2012review}
Ken Brodlie, Rodolfo~Allendes Osorio, and Adriano Lopes.
\newblock A review of uncertainty in data visualization.
\newblock {\em Expanding the frontiers of visual analytics and visualization},
  pages 81--109, 2012.

\bibitem{browne2015dark}
Simone Browne.
\newblock {\em Dark matters}.
\newblock Duke University Press, 2015.

\bibitem{chiou2021trusting}
Erin~K Chiou and John~D Lee.
\newblock Trusting automation: Designing for responsivity and resilience.
\newblock {\em Human Factors}, page 00187208211009995, 2021.

\bibitem{crawford2021atlas}
Kate Crawford.
\newblock {\em The Atlas of AI}.
\newblock Yale University Press, 2021.

\bibitem{dietvorst2015algorithm}
Berkeley~J Dietvorst, Joseph~P Simmons, and Cade Massey.
\newblock Algorithm aversion: People erroneously avoid algorithms after seeing
  them err.
\newblock {\em Journal of Experimental Psychology: General}, 144(1):114, 2015.

\bibitem{dovsilovic2018explainable}
Filip~Karlo Do{\v{s}}ilovi{\'c}, Mario Br{\v{c}}i{\'c}, and Nikica Hlupi{\'c}.
\newblock Explainable artificial intelligence: A survey.
\newblock In {\em 2018 41st International convention on information and
  communication technology, electronics and microelectronics (MIPRO)}, pages
  0210--0215. IEEE, 2018.

\bibitem{eubanks2018automating}
Virginia Eubanks.
\newblock {\em Automating Inequality: How High-tech Tools Profile, Police, and
  Punish the Poor}.
\newblock St. Martin's Press, 2018.

\bibitem{ferreira2020people}
Juliana~J Ferreira and Mateus~S Monteiro.
\newblock What are people doing about xai user experience? a survey on ai
  explainability research and practice.
\newblock In {\em International Conference on Human-Computer Interaction},
  pages 56--73. Springer, 2020.

\bibitem{griethe2006visualization}
Henning Griethe, Heidrun Schumann, et~al.
\newblock The visualization of uncertain data: Methods and problems.
\newblock In {\em SimVis}, pages 143--156, 2006.

\bibitem{hameleers2020picture}
Michael Hameleers, Thomas~E Powell, Toni~GLA Van Der~Meer, and Lieke Bos.
\newblock A picture paints a thousand lies? the effects and mechanisms of
  multimodal disinformation and rebuttals disseminated via social media.
\newblock {\em Political Communication}, 37(2):281--301, 2020.

\bibitem{han2020beyond}
Wenkai Han and Hans-J{\"o}rg Schulz.
\newblock Beyond trust building—calibrating trust in visual analytics.
\newblock In {\em 2020 IEEE workshop on trust and expertise in visual analytics
  (TREX)}, pages 9--15. IEEE, 2020.

\bibitem{heylighendewaele1999formality}
Francis Heylighen and Jean-Marc Dewaele.
\newblock Formality of language: definition, measurement and behavioral
  determinants.
\newblock {\em Interner Bericht, Center “Leo Apostel”, Vrije Universiteit
  Br{\"u}ssel}, 4, 1999.

\bibitem{hullman2018pursuit}
Jessica Hullman, Xiaoli Qiao, Michael Correll, Alex Kale, and Matthew Kay.
\newblock In pursuit of error: A survey of uncertainty visualization
  evaluation.
\newblock {\em IEEE transactions on visualization and computer graphics},
  25(1):903--913, 2018.

\bibitem{kelleher2009conversational}
Tom Kelleher.
\newblock Conversational voice, communicated commitment, and public relations
  outcomes in interactive online communication.
\newblock {\em Journal of communication}, 59(1):172--188, 2009.

\bibitem{Lazer1094}
David M.~J. Lazer, Matthew~A. Baum, Yochai Benkler, Adam~J. Berinsky, Kelly~M.
  Greenhill, Filippo Menczer, Miriam~J. Metzger, Brendan Nyhan, Gordon
  Pennycook, David Rothschild, Michael Schudson, Steven~A. Sloman, Cass~R.
  Sunstein, Emily~A. Thorson, Duncan~J. Watts, and Jonathan~L. Zittrain.
\newblock The science of fake news.
\newblock {\em Science}, 359(6380):1094--1096, 2018.

\bibitem{mayr2019trust}
Eva Mayr, Nicole Hynek, Saminu Salisu, and Florian Windhager.
\newblock Trust in information visualization.
\newblock In {\em TrustVis@ EuroVis}, pages 25--29, 2019.

\bibitem{nass2005wired}
Clifford~Ivar Nass and Scott Brave.
\newblock {\em Wired for speech: How voice activates and advances the
  human-computer relationship}.
\newblock MIT Press, Cambridge,MA, 2005.

\bibitem{pennycook2020fighting}
Gordon Pennycook, Jonathon McPhetres, Yunhao Zhang, and David Rand.
\newblock Fighting covid-19 misinformation on social media: Experimental
  evidence for a scalable accuracy nudge intervention.
\newblock {\em Psychological Science}, 0(0), 2020.

\bibitem{reeves1996media}
Byron Reeves and Clifford Nass.
\newblock {\em The media equation: How people treat computers, television, and
  new media like real people}.
\newblock Cambridge University Press, Cambridge, UK, 1996.

\bibitem{sacha2015role}
Dominik Sacha, Hansi Senaratne, Bum~Chul Kwon, Geoffrey Ellis, and Daniel~A
  Keim.
\newblock The role of uncertainty, awareness, and trust in visual analytics.
\newblock {\em IEEE transactions on visualization and computer graphics},
  22(1):240--249, 2015.

\bibitem{schwarz2016making}
Norbert Schwarz, Eryn Newman, and William Leach.
\newblock Making the truth stick \& the myths fade: Lessons from cognitive
  psychology.
\newblock {\em Behavioral Science \& Policy}, 2(1):85--95, 2016.

\bibitem{sinha2020unnatural}
Koustuv Sinha, Prasanna Parthasarathi, Joelle Pineau, and Adina Williams.
\newblock {UnNatural} {L}anguage {I}nference.
\newblock In {\em Proceedings of the 59th Annual Meeting of the Association for
  Computational Linguistics and the 11th International Joint Conference on
  Natural Language Processing (Volume 1: Long Papers)}, pages 7329--7346,
  Online, August 2021. Association for Computational Linguistics.

\bibitem{xu2019explainable}
Feiyu Xu, Hans Uszkoreit, Yangzhou Du, Wei Fan, Dongyan Zhao, and Jun Zhu.
\newblock Explainable ai: A brief survey on history, research areas, approaches
  and challenges.
\newblock In {\em CCF international conference on natural language processing
  and Chinese computing}, pages 563--574. Springer, 2019.

\end{thebibliography}

\end{document}